\documentclass[cameraready]{Interspeech}

\title{Self-Supervised Speech Models Encode Phonetic Context via Position-dependent Orthogonal Subspaces}

\author[affiliation={1}]{Kwanghee}{Choi}
\author[affiliation={1}]{Eunjung}{Yeo}
\author[affiliation={2}]{Cheol Jun}{Cho}
\author[affiliation={3}]{David R.}{Mortensen}
\author[affiliation={1}]{David}{Harwath}

\address{
    $^1$ UT Austin \quad
    $^2$ UC Berkeley \quad
    $^3$ Carnegie Mellon University
}

\email{\{kwanghee,harwath\}@utexas.edu}


\usepackage{cite}
\usepackage{cleveref}
\usepackage{tipa}
\usepackage{graphicx}
\usepackage{subcaption}

\crefname{section}{\S}{\S\S}
\Crefname{section}{\S}{\S\S}
\crefname{appendix}{\S}{\S\S}
\Crefname{appendix}{\S}{\S\S}

\begin{document}

\maketitle

\begin{abstract}
Transformer-based self-supervised speech models (S3Ms) are often described as contextualized, yet what this entails remains unclear.
Here, we focus on how a single frame-level S3M representation can encode phones and their surrounding context.
Prior work has shown that S3Ms represent phones compositionally; for example, phonological vectors such as voicing, bilabiality, and nasality vectors are superposed in the S3M representation of [m].
We extend this view by proposing that phonological information from a sequence of neighboring phones is also compositionally encoded in a single frame, such that vectors corresponding to previous, current, and next phones are superposed within a single frame-level representation.
We show that this structure has several properties, including orthogonality between relative positions, and emergence of implicit phonetic boundaries.
Together, our findings advance our understanding of context-dependent S3M representations.\footnote{All code for experiments is available at 
\ifcameraready
\url{https://github.com/juice500ml/phonetic-arithmetic}.
\else
\url{https://anonymous.4open.science/r/phonetic-arithmetic-1CF2}.
\fi
}
\end{abstract}

\section{Introduction}
Self-supervised speech models (S3Ms) have become a central paradigm for learning general-purpose speech representations from unlabeled audio, enabling strong performance across a wide range of downstream tasks \cite{baevski2020wav2vec,hsu2021hubert,chen2022wavlm}.
The neural architecture of this class of models was introduced by wav2vec \cite{schneider2019wav2vec} and has since been widely adopted by subsequent S3Ms \cite{baevski2020wav2vec,hsu2021hubert,chen2022wavlm}.
The architecture consists of two main components: a convolutional network that transforms the raw waveform into a sequence of latent representations, and a transformer-based context network that can combine information across time.

While this architecture has proven highly effective, its internal mechanisms remain only partially understood.
Prior work has identified \textit{what} kinds of acoustic and linguistic information is present in different layers of S3Ms \cite{pasad2021layer,pasad2023comparative}.
However, despite covering multiple levels of the linguistic hierarchy, it remains unclear \textit{how} the context network organizes such information within the representation space.
This gap has motivated a growing body of work aimed at characterizing the internal structure of representations learned by S3Ms \cite{liu2023self,meng25b_interspeech,choi2022opening,choi2024self,choi2026self,choi2025leveraging,sicherman2023analysing,wells22_interspeech}.

\begin{figure}[t!]
    \centering
    \begin{subfigure}[t]{\linewidth}
        \centering
        \includegraphics[width=\linewidth]{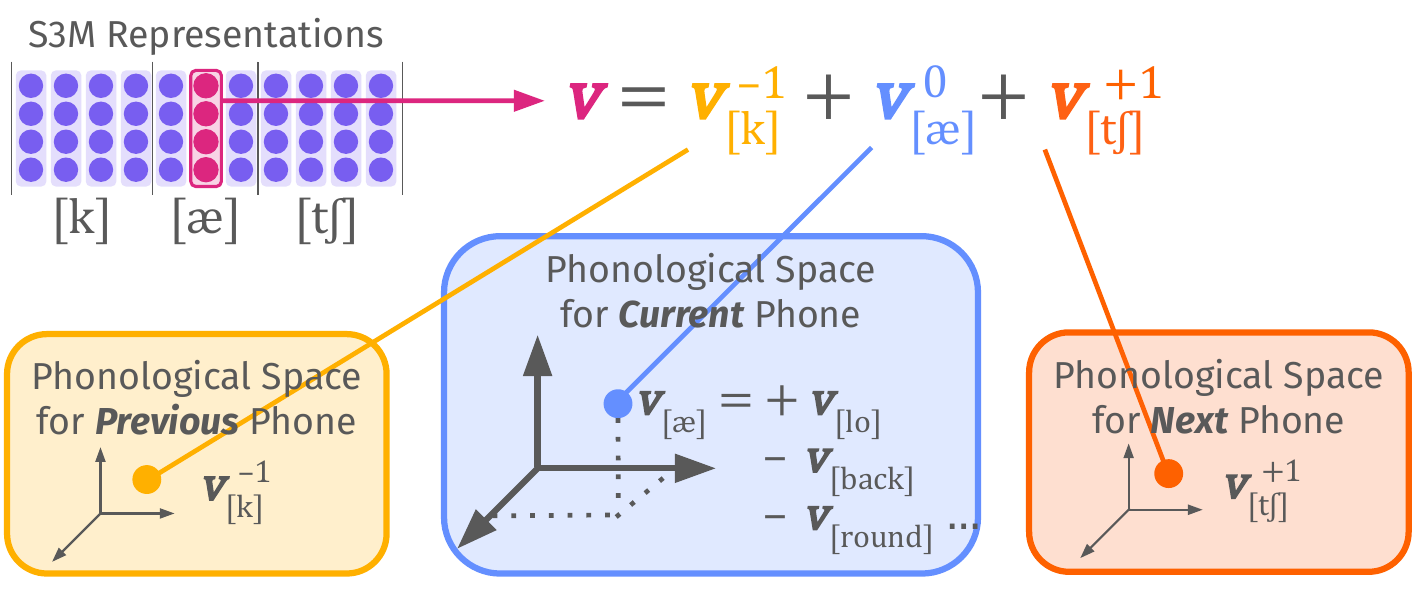}
        \caption{
            Conceptual illustration of our hypothesis.
        }
        \label{fig:summary}
    \end{subfigure}

    \vspace{1em}

    \begin{subfigure}[t]{\linewidth}
        \centering
        \includegraphics[width=\linewidth]{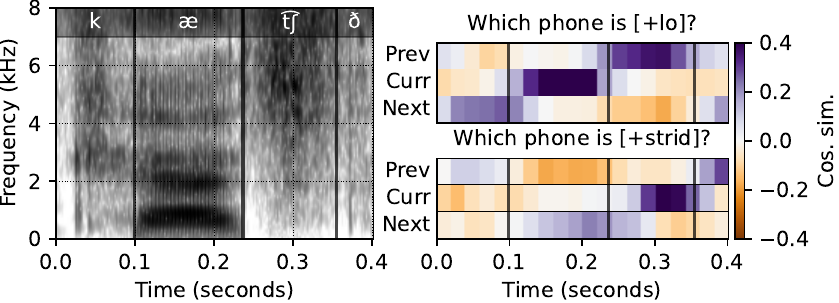}
        \caption{
            Empirical demonstration of our hypothesis.
        }
        \label{fig:evidence}
    \end{subfigure}
    \caption{
        (a) Each frame-level S3M representation contains phonological vectors from position-sensitive, orthogonal subspaces.
        (b) Cosine similarity between frame-level representations and phonological vectors (\textit{low}, \textit{strident}) from the previous, current, and next phone subspaces. 
        Individual frames encode phonological information from the current and neighboring phones with respect to phonetic boundaries.
    }
    \label{fig:fig1}
\end{figure}


\cite{pasad2021layer} has shown that the convolutional networks of S3Ms primarily capture local acoustic information.
Similar to a spectrogram, the limited receptive field enforces temporal locality, with each frame aligned to a specific region of the input signal.
Further, \cite{choi2022opening} showed that, unlike spectrograms---whose dimensions correspond to fixed frequency bands---the convolutional representation space is distributed and relative.
In this space, acoustic similarity is reflected in the cosine similarity between representations rather than along fixed axes.
These findings suggest that while convolutional network can provide local acoustic information, they do not by themselves account for how information beyond local acoustics is incorporated within S3Ms.

Despite this relatively clear picture concerning the convolutional network, the role of the transformer-based context network is less well understood.
Empirically, the context network is crucial.
Representations from later transformer layers consistently yield stronger performance on speech probing tasks, such as phone and word identity, compared to convolutional representations \cite{pasad2023comparative}, and combining information across transformer layers often further improves performance \cite{shih2024interface}.
Moreover, scaling the context network typically leads to substantial gains, with \texttt{Large} models consistently outperforming \texttt{Base} models \cite{baevski2020wav2vec,hsu2021hubert,chen2022wavlm}.
These trends suggest that the context network contributes more than a simple refinement of local acoustics.

An additional clue comes from the effectiveness of discretized S3M representations in various speech applications \cite{chang2024exploring}, where k-means clustering of S3M representations yields discrete units that align with phonetic categories despite the absence of supervision \cite{wells22_interspeech,baevski2020wav2vec,hsu2021hubert,de2024human}.
Moreover, the units tend to organize according to phonetic categories.
For example, phones from the same category exhibit higher mutual similarity than phones from different category (\textit{e.g.}, vowels vs. consonants) \cite{abdullah2023information,sicherman2023analysing}.
Finally, phonetic and syllabic boundaries have also been observed in attention maps \cite{li2023dissecting}, enabling the use of S3Ms for unsupervised speech segmentation \cite{pasad2024self,baade2025syllablelm,cho2025sylber,cho2026sylber,visser2026zerosyl}.
These observations suggest that context networks encode linguistically meaningful distinctions rather than merely reflecting fine-grained acoustic similarity.

To explain the emergence of such linguistically structured representations, recent work has proposed that S3Ms encode phonological vectors \cite{choi2026self}.
Under this view, individual phones are represented with linear combinations of phonological vectors, yielding a compositional representation space.
For example, the S3M representation of \textipa{[\ae]} contains a linear combination of phonological vectors, including low, front, and unrounded vectors, in a way that supports vector arithmetic.
While this framework provides a coherent explanation for phonological capabilities described above, it considers only individual phones in isolation and does not examine how phonetic context from neighboring phones are captured by transformers.

This motivates us to extend their framework by hypothesizing that \textit{each frame-level S3M representation encodes not only a single phone, but sequences of phones} (\Cref{fig:fig1}).
This extension aims to shed light on how S3Ms encode contextual information.
Specifically, we propose each contextualized frame-level representation contains a linear combination of phonological vectors corresponding not only to the phone aligned with that frame, but also to its neighboring phones.
Based on this hypothesis, we formulate the following four predictions.

\textbf{Prediction 1 (\Cref{ss:pooling}).}
Individual frame-level representations will exhibit a compositional phonological structure.
We assess whether vector arithmetic within the representation space reflects phonological analogies, \textit{e.g.}, adding a voicing vector \textipa{[d]} $-$ \textipa{[t]} to a voiceless stop \textipa{[p]} approximates the voiced stop \textipa{[b]}.
While \cite{choi2026self} rely on mean-pooled representations, we evaluate whether such phonological analogies hold at the level of individual frame representations.

\textbf{Prediction 2 (\Cref{ss:neighbors}).}
Individual frame-level representations will encode phonological information of neighboring phone segments as well.
To test this, we extend the above phonological analogy test to neighboring phones, examining whether individual frame representations support phonological analogies on previous or next phones.
Concretely, we test whether there exist vectors in the representation space that encode contextual phonological properties such as ``is the previous phone voiced?'' or ``is the next phone a high vowel?'', and whether such properties can be recovered via linear operations on single-frame representations.

\textbf{Prediction 3 (\Cref{ss:orthogonality}).}
Position-dependent subspaces will be orthogonal to one another, allowing the model to distinguish which information corresponds to each relative position, \textit{e.g.}, previous, current, or next phone.
For example, if the current phone and the previous phone are voiced, a representation that simply sums shared phonological vectors cannot distinguish whether the voicing originated from the current or previous phone.
To resolve such collisions, contextualized representations should encode phonological information in a position-dependent manner, with phonological vectors associated with different relative positions occupying distinct subspaces.
We therefore test whether phonological vectors associated with different relative positions are orthogonally encoded in the single-frame representations.

\textbf{Prediction 4 (\Cref{ss:segmentation}).}
Phonetic boundaries will be implied by temporal changes of frame-level representations due to underlying position-dependent subspaces.
To distinguish among the previous, current, and next phones, the model may implicitly infer phonetic boundaries and segment its frame-level representations accordingly.
One other possibility is that position-dependent phonological information ignores phonetic boundaries and instead operates over a fixed symmetric window \cite{meng25b_interspeech,choi2025device}.
To distinguish between these possibilities, we examine how the cosine similarity between frame-level representations and position-dependent phonological subspaces changes across the frames, particularly around phonetic boundaries.

To summarize, we aim to explain how S3Ms encode phonetic context, \textit{i.e.}, sequence of phones rather than individual phones.
We hypothesize that frame-level S3M representations encode phonological information in a compositional and position-sensitive manner, and test the following predictions:
\begin{itemize}
    \item \textbf{Frame-level compositionality}: individual frame representations can be expressed as linear combinations of phonological vectors describing the phone aligned with that frame (\Cref{ss:pooling}).
    \item \textbf{Contextual phonological vectors}: individual frame representations encode phonological information about neighboring phones (\Cref{ss:neighbors}).
    \item \textbf{Positional orthogonality}: phonological vectors associated with relative positions (previous, current, and next phones) occupy mutually orthogonal subspaces (\Cref{ss:orthogonality}).
    \item \textbf{Phonetic segmentation}: position-dependent phonological subspaces within frame-level representations are determined based on phonetic boundaries (\Cref{ss:segmentation}).
\end{itemize}

\section{Settings}
\subsection{Self-supervised Speech Models (S3Ms)}
We analyze representations from three widely used monolingual S3Ms trained on English: wav2vec 2.0 \cite{baevski2020wav2vec}, HuBERT \cite{hsu2021hubert}, and WavLM \cite{chen2022wavlm}, using the publicly released \texttt{Large} checkpoints.\footnote{We used the checkpoints \texttt{facebook/wav2vec2-large-lv60}, \texttt{facebook/hubert-large-ll60k}, and \\ \texttt{microsoft/wavlm-large}.}
wav2vec 2.0 and HuBERT are pretrained primarily on audiobooks, whereas WavLM uses both read and spontaneous speech.
All three models share a common architecture consisting of a convolutional feature encoder followed by a transformer-based context network.

Following \cite{choi2026self}, we use spectral representations, log-mel spectrograms (MelSpec) and MFCCs, as baselines.
We use the \texttt{librosa} implementation \cite{mcfee2015librosa}, with a window length of 2048, a hop length of 512, and without delta and delta-delta coefficients for MFCCs.
Following \cite{pasad2021layer,pasad2023comparative}, we extract phone-level representations using mean pooling over frames within each phone segment, and apply the same pooling procedure to the spectral and S3M representations as baselines for \Cref{ss:pooling,ss:neighbors}.

\subsection{S3M Analysis via Phonological Analogies}
\textbf{Phonological analogies.}
Phonological analogy-based analysis \cite{chaabouni2017learning,zouhar2024pwesuite,choi2026self} test whether phonological relationships correspond to linear operations in the representation space.
Given a quadruplet of phones forming a consistent phonological contrast, the analysis evaluates whether the difference between two phones (\textit{e.g.}, \textipa{[b]} $-$ \textipa{[d]}) can be added to another phone (\textit{e.g.}, \textipa{[t]}) to approximate the expected target (\textit{e.g.}, \textipa{[p]}), such that shared phonological differences are reflected by vector subtractions.
Phonological analogies are automatically created using \texttt{PanPhon} \cite{mortensen2016panphon} and phonological contrasts may involve multiple \texttt{PanPhon} feature differences.
For example, \textipa{[b]} and \textipa{[t]} differs by \texttt{[voi]}, \texttt{[cor]}, \texttt{[distr]}, and \texttt{[lab]}.

\textbf{Success rate.}
To quantify whether such phonological analogies hold, we follow the success rate metric of \cite{choi2026self}.
For each analogy, we apply vector arithmetic in order to approximate the representation of a target phone (\textit{e.g.}, approximating \textipa{[p]} using \textipa{[b]} $-$ \textipa{[d]} $+$ \textipa{[t]}).
We then compare the cosine similarity between the target and the approximated representation (\textit{e.g.}, $\cos$(\textipa{[p]}, approximated \textipa{[p]}) to upper and lower baselines.
The upper baseline is given by the similarity between same phone across different utterances (\textit{e.g.}, $\cos$(\textipa{[p]}, \textipa{[p]} from different utterance)).
The lower baseline given by the similarity between different phones (\textit{e.g.}, $\cos$(\textipa{[p]}, any phone that is not \textipa{[p]})).
An analogy is considered successful if the averaged cosine similarity falls between these baselines, and the success rate is defined as the proportion of analogies satisfying this ordering.
For experiments in \Cref{ss:pooling,ss:neighbors}, we adopt the same hyperparameters as \cite{choi2026self}: 1{,}000 random samples drawn with replacement to estimate average cosine similarity, and a 99\% confidence interval computed from 10 independent replications.

\subsection{Datasets}
We conduct experiments on TIMIT \cite{garofolo1993darpa} and VoxAngeles \cite{chodroff2024voxangeles}, phonetically transcribed and manually segmented speech datasets.
TIMIT consists of sentence-level read English speech, while VoxAngeles contains word-level read speech from 95 languages unseen during S3M training.
Following \cite{choi2026self}, we exclude phones without a \texttt{PanPhon} mapping, reducing the inventories from 47 to 47 to 44 phones for TIMIT and from 567 to 468 phones for VoxAngeles.
These filtered sets are used in \Cref{ss:orthogonality,ss:segmentation,ss:demo}.
Also, to ensure reliable success rate calculation, we further exclude phones that occur fewer than 50 times, resulting in 43 phones for TIMIT and 57 for VoxAngeles.
This yields 236 and 468 valid quadruplets, respectively, used in \Cref{ss:pooling,ss:neighbors}.
Unless stated otherwise, we use the TIMIT test split and the full VoxAngeles dataset.

\section{Experiments}

\subsection{Frame-level compositionality}\label{ss:pooling}
\textbf{Settings.}
To assess whether phonological compositionality is present at the frame level, we compare mean pooling \cite{pasad2021layer,pasad2023comparative,choi2026self} to center pooling \cite{pasad2024self,choi2024self,choi2025leveraging}.
A phone segment typically spans multiple frames.
Mean pooling aggregates frame-level representations across the segment, whereas center pooling uses only the temporally central frame.
We assess whether a single center frame, compared to mean pooling, can also capture the phonological information.

\begin{figure}[t]
    \centering
    \includegraphics[width=0.9\columnwidth]{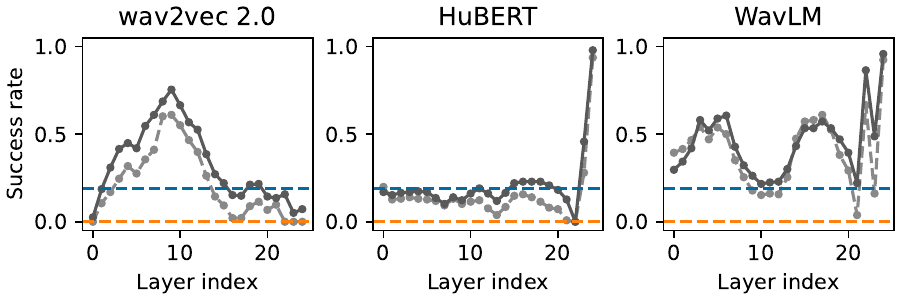}
    \includegraphics[width=0.9\columnwidth]{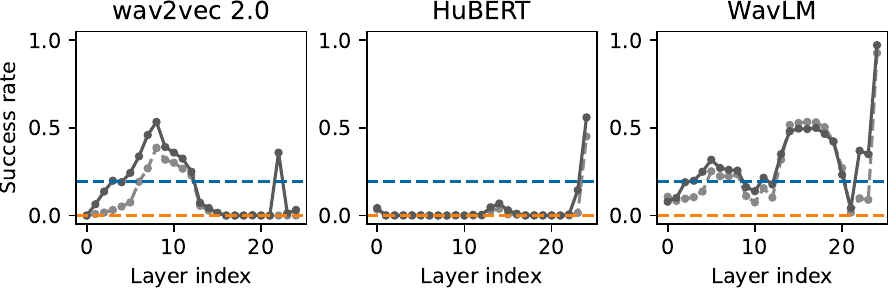}
    \includegraphics[width=\columnwidth]{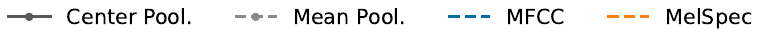}
    \caption{
        Phonological analogy success rates comparing mean and center pooling on TIMIT (above) and VoxAngeles (below).
        Center pooling is on par and often outperforms mean-pooling, indicating that phonological compositionality is present at the level of individual frame representations.
    }
    \label{fig:pooling}
\end{figure}

\textbf{Results.}
As shown in \Cref{fig:pooling}, center-pooled representations show comparable and often higher success rates than mean-pooled representations across datasets and S3Ms, and both substantially outperform spectral baselines in certain layers.
This result indicates that phonological compositionality does not rely on temporal averaging across a phone segment, \textit{i.e.}, mean pooling.
Instead, a single frame-level representation supports linear phonological analogies, consistent with our hypothesis of frame-level phonological compositionality.

\subsection{Contextual phonological vectors}\label{ss:neighbors}
\subsubsection{How much neighboring context does the center frame encode?}\label{ss:neighbors-center}
\textbf{Settings.}
A natural extension of the previous analysis is to ask how much neighboring phones' phonological information is encoded in a single frame-level representation.
Specifically, we investigate whether a center-pooled representation supports phonological analogies for not only about the current phone, but also about neighboring phones.

Consider a sequence of phones $[p^{-2}, p^{-1}, p^{0}, p^{+1}, p^{+2}]$.
For each sequence, we extract a single center-pooled representation from the center phone $p^{0}$ and use this representation to probe phonological information associated with each relative position.
This allows us to test whether the center frame of $p^{0}$ encodes phonological vectors corresponding to phones beyond $p^0$.
For example, to test whether the center representation from $p^{0}$ contains the voicing vector of the previous phone $p^{-1}$, we compare representations from contexts in which $p^{-1}$ is voiced or voiceless phones.

\begin{figure}[t!]
    \centering
    \includegraphics[width=0.9\columnwidth]{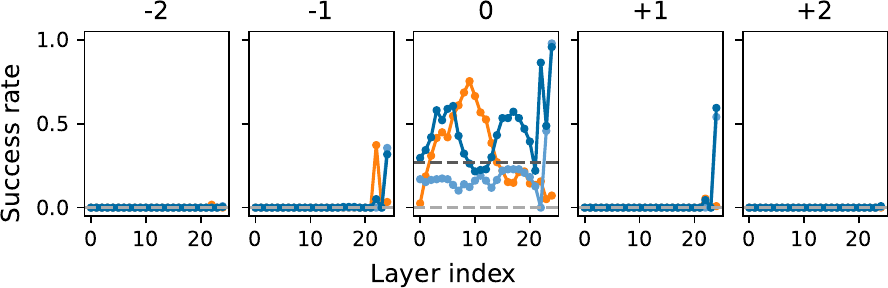}
    \includegraphics[width=0.9\columnwidth]{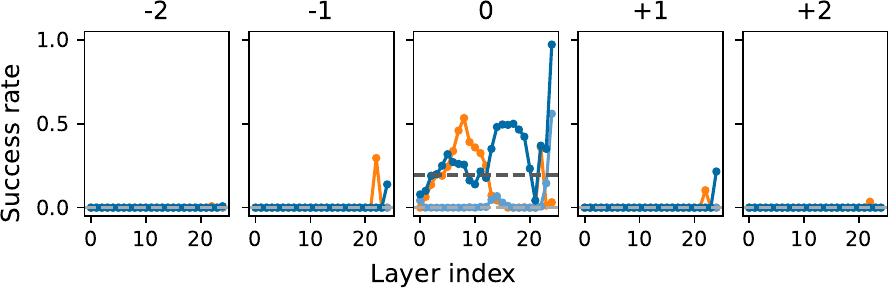}
    \includegraphics[width=\columnwidth]{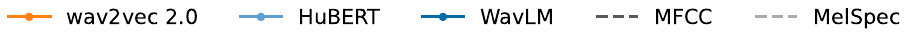}
    \caption{
        Phonological analogy success rates for probing contextual information encoded in a single frame-level representation of $p^{0}$ on TIMIT (upper) and VoxAngeles (lower).
        Center-pooled S3M representations support phonological analogies for the current phone position $(0)$ and its neighbors $(\pm 1)$.
    }
    \label{fig:neighbors}
\end{figure}

\textbf{Results.}
As shown in \Cref{fig:neighbors}, center-pooled S3M representations support phonological analogies not only for the current phone, but also for its immediate neighbors, $p^{-1}$, $p^{0}$, and $p^{+1}$.
Specifically, nonzero success rate is observed particularly in the later layers of S3Ms, indicating that a single frame-level representation from the center phone encodes phonological information from neighboring phones.

\subsubsection{Effective window size for phonological analogies}\label{ss:neighbors-random}
\textbf{Settings.}
To further localize where the phonological information of the center phone $p^{0}$ is encoded, we evaluate representations extracted from frames at different temporal positions within the phones $p^{-2}$ through $p^{+2}$.
Given frame-level representations $\mathbf{r}[1], \mathbf{r}[2], \ldots, \mathbf{r}[n]$ aligned to a phone segment of length $n$ frames (where the segment may come from any phone from $p^{-2}$ to $p^{+2}$), we perform \emph{random pooling}:
\begin{align}
    \mathbf{r}^\text{random} &= \mathbf{r}[i]\text{ where } i \sim \text{Unif}\{1, \dots, n\}.
\end{align}
We bin representations by their normalized position $i/n$ within each phone segment and aggregate results across bins.
For VoxAngeles, unlike TIMIT, binning by relative position reduces the number of usable phonological quadruplets, from 468 to 128.
Based on \Cref{ss:neighbors-center}, we choose the S3M layer that exhibited the strongest performance for $p^{0}$: the 24th layer for HuBERT and WavLM, and the 9th layer for wav2vec~2.0.

\begin{figure}[t!]
    \centering
    \includegraphics[width=1.0\columnwidth]{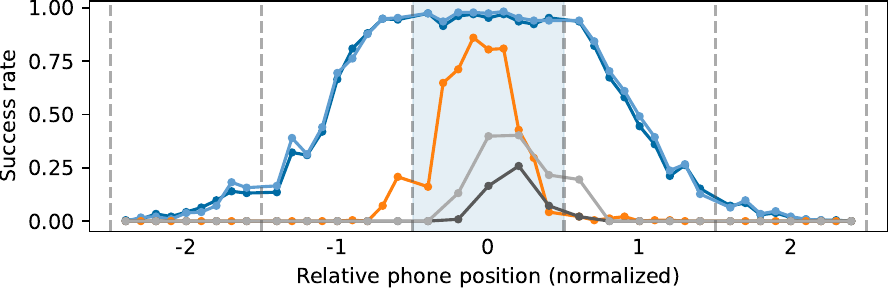}
    \includegraphics[width=1.0\columnwidth]{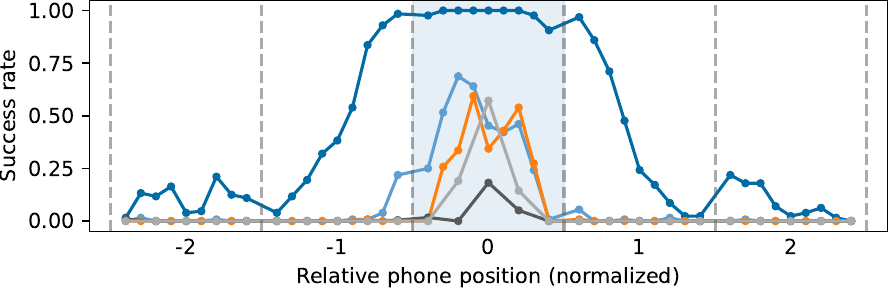}
    \includegraphics[width=\columnwidth]{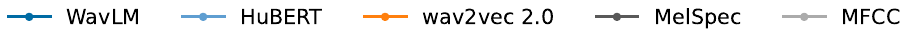}
    \caption{
        Phonological analogy success rate for center phone $p^{0}$ with respect to relative position on TIMIT (upper) and VoxAngeles (lower).
        WavLM exhibits the widest window, with high success rates within the center phone position ${0}$, decreasing for $\pm1$, and near-zero for $\pm2$.
        Spectral representations, unlike S3Ms, show nonzero success rates only near the center.
    }
    \label{fig:random}
\end{figure}

\textbf{Results.}
\Cref{fig:random} illustrates the effective window over which frame-level representations support phonological analogies for the center phone $p^{0}$.
Among the evaluated representations, WavLM exhibits the clearest trapezoidal pattern across datasets.
For position $0$, the success rate remains consistently high; it decreases for the immediately neighboring positions $\pm 1$ and approaches zero for more distant positions $\pm 2$.

HuBERT exhibits a similar pattern on TIMIT, but this behavior weakens on VoxAngeles.
In contrast, wav2vec 2.0 demonstrates limited performance on both datasets.
Finally, as expected, the spectral baselines are effective only at position 0.

Through the lens of coarticulation, S3M representations may integrate phonological information from neighboring phones in a way that supports reconstruction of the center phone, where \Cref{ss:reconstruction} discusses in detail.
We further observe that the success rate seems to vary with respect to phonetic boundaries, which we analyze more directly in \Cref{ss:segmentation}.

\subsection{Positional orthogonality} \label{ss:orthogonality}
\textbf{Settings.}
Results in \Cref{ss:neighbors} show that a single frame-level representation in the current phone $p^0$ can encode phonological information about neighboring phones $p^{\pm1}$.
This creates an ambiguity: if phonological features from the current and neighboring phones are the same, the representation must distinguish their positional sources.
We hypothesize that phonological information associated with different relative positions is encoded in orthogonal subspaces.

To test this hypothesis, we follow \cite{choi2026self} and extract \emph{phonological vectors} via difference-of-means.
For example, the voicing vector associated with the current phone is computed as:
\begin{align}
    \mathbf{v}^{0}_{\text{voicing}} 
    = \mathbb{E}_{p^{0}\,\text{is voiced}}[\mathbf{v}]
    - \mathbb{E}_{p^{0}\,\text{is unvoiced}}[\mathbf{v}],
\end{align}
where $\mathbf{v}$ denotes a frame-level representation obtained via center pooling.
Rather than restricting to minimal pairs, we average representations across all phones that do or do not have the target phonological feature.
Analogously, we extract a voicing vector for the previous phone $p^{-1}$:
\begin{align}
    \mathbf{v}^{-1}_{\text{voicing}} 
    = \mathbb{E}_{p^{-1}\,\text{is voiced}}[\mathbf{v}]
    - \mathbb{E}_{p^{-1}\,\text{is unvoiced}}[\mathbf{v}].
\end{align}
If positional orthogonality holds, these vectors should be approximately orthogonal, i.e.,
$\cos(\mathbf{v}^{0}_{\text{voi.}}, \mathbf{v}^{-1}_{\text{voi.}}) \approx 0$,
allowing a single frame-level representation to encode phonological information from multiple positions without confusion.

Following \cite{choi2026self}, we compute eight phonological vectors from four vowel features (high, low, back, round) and four consonantal features (nasal, sonorant, strident, voicing).
For vowel features, phonological vectors are extracted using vowels only, and for consonantal features, vectors are extracted using consonants only.
All analyses use the final layer of WavLM-Large, which consistently achieved the strongest performance in \Cref{ss:neighbors}.
We use the training split of the datasets to compute the phonological vectors.
For VoxAngeles, the training split is created by randomly selecting a subset of languages from the dataset.

\begin{figure}[t]
    \centering
    \includegraphics[width=0.9\columnwidth]{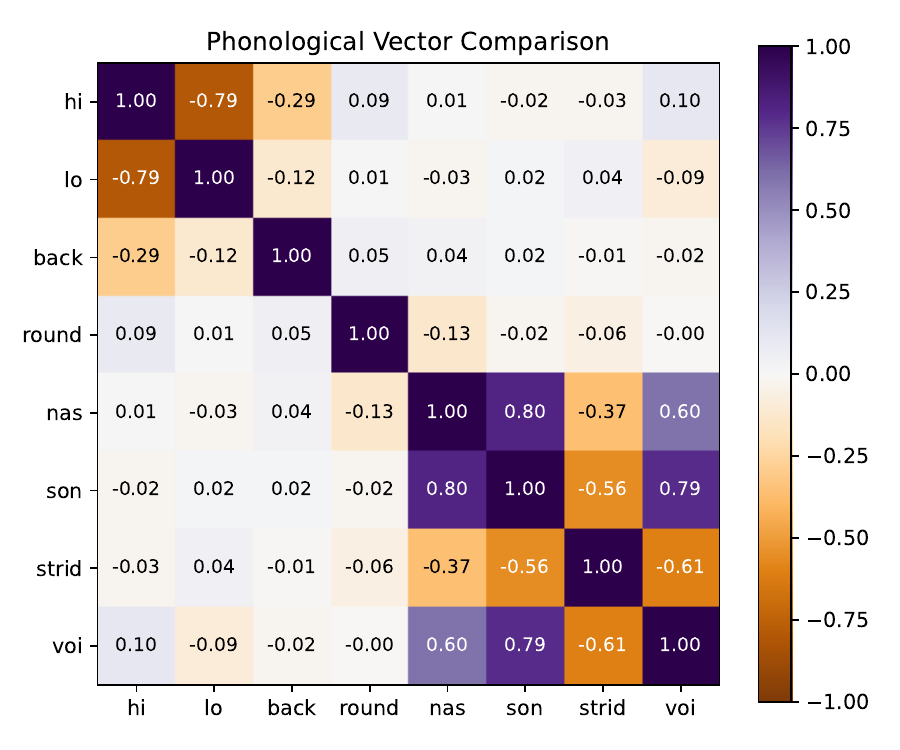}
    \includegraphics[width=0.9\columnwidth]{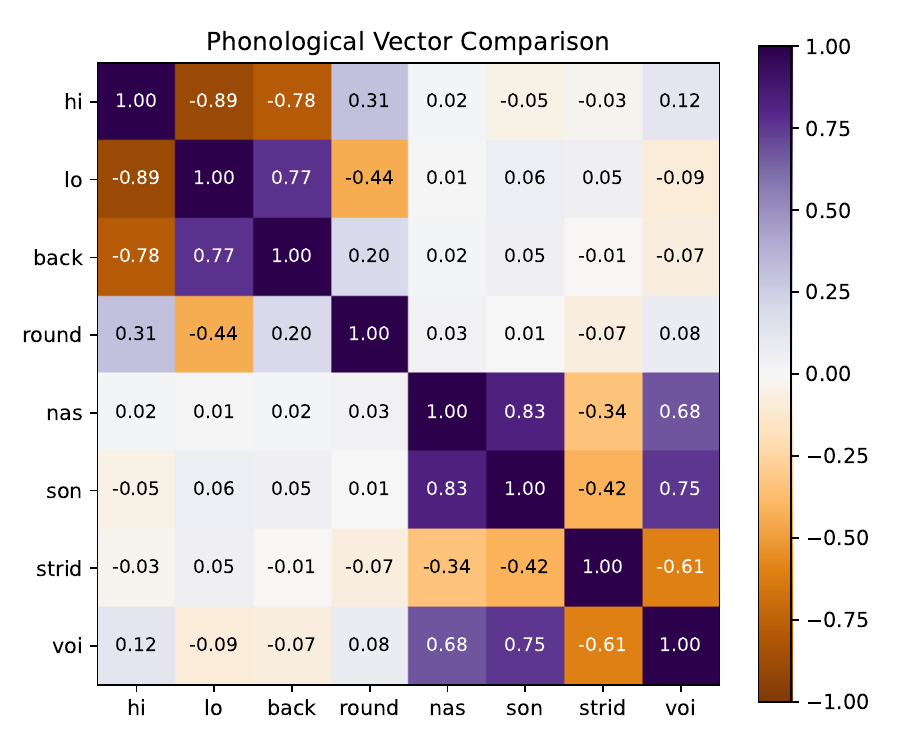}
    \caption{
        Cosine similarity between phonological vectors extracted from frame-level S3M representations on TIMIT (upper) and VoxAngeles (lower).
        The structure mirrors that of \cite{choi2026self}, with opposing features showing negative similarity and related features showing positive similarity.
    }
    \label{fig:phonovectors}
\end{figure}

\begin{figure}[t]
    \centering
    \includegraphics[width=0.90\columnwidth]{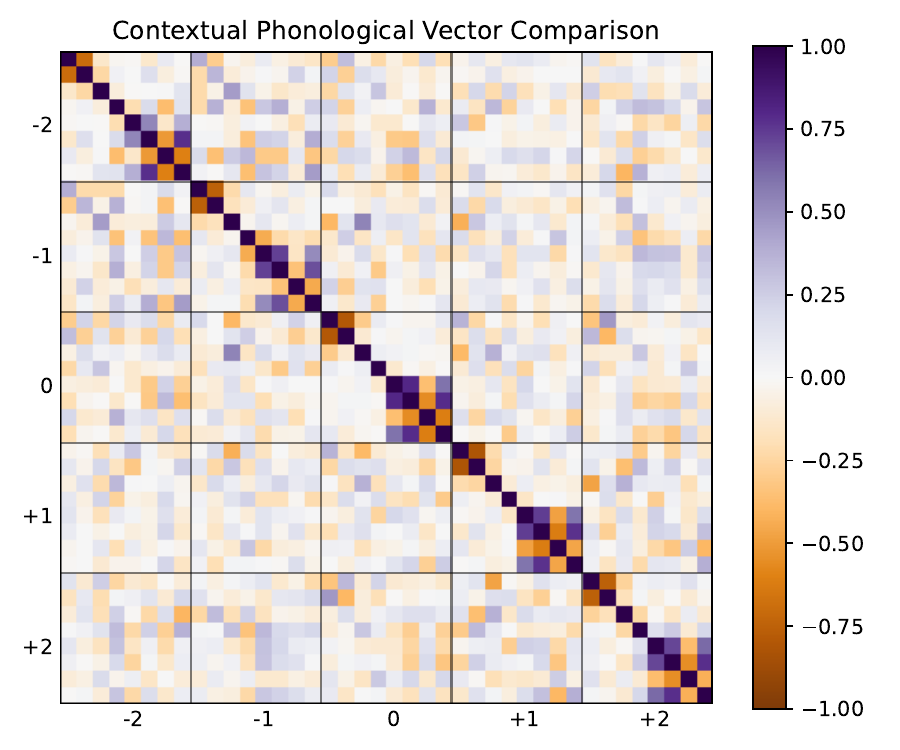}
    \includegraphics[width=0.9\columnwidth]{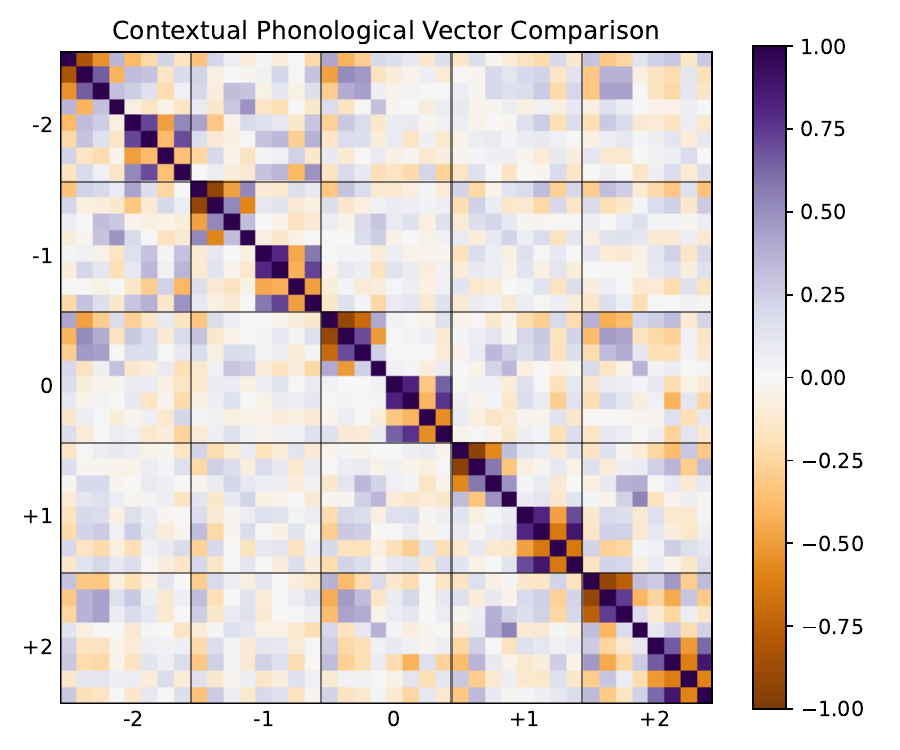}
    \caption{
        Cosine similarity between phonological vectors associated with different relative phone positions ($-2$ to $+2$) from TIMIT (upper) and VoxAngeles (lower).
        Comparing with \Cref{fig:phonovectors}, relative similarity structure is preserved within positions.
        Further, vectors from different positions exhibit substantially lower similarity than those from the same position, implying approximate positional orthogonality.
    }
    \label{fig:orthogonality}
\end{figure}

\begin{figure}[t!]
    \centering
    \includegraphics[width=0.95\columnwidth]{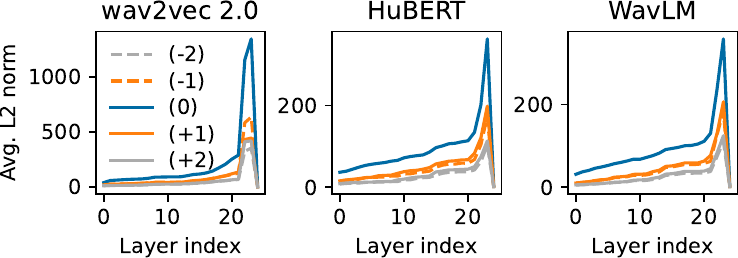}
    \includegraphics[width=0.95\columnwidth]{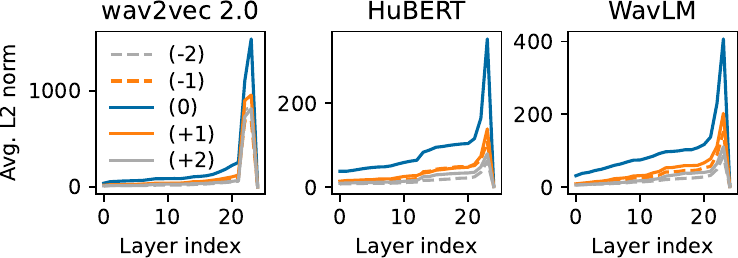}
    \caption{
    Average L2 norm of eight phonological vectors per its relative position for each layer on TIMIT (upper) and VoxAngeles (lower).
    We observe that the average vector size decreases with increasing distance from the center phone.
    }
    \label{fig:layerwise_norm}
\end{figure}

\textbf{Results.}
We first examine cosine similarities among phonological vectors extracted from the current phone using center-pooled frame-level representations (\Cref{fig:phonovectors}).
The resulting similarity structure closely matches that reported in \cite{choi2026self} that used mean pooling.
Notably, the structure aligns well with linguistic intuition.
For example, opposing features such as \textit{high} and \textit{low} exhibit strong negative similarity, while related features such as \textit{nasal}, \textit{sonorant}, and \textit{voicing} show positive similarity.
In addition, vowel and consonant features are approximately orthogonal.

Next, we compare phonological vectors across relative positions (\Cref{fig:orthogonality}).
Phonological vectors associated with different relative positions exhibit substantially lower cosine similarity with one another than with vectors from the same position, showing approximate orthogonality across positions.
Additionally, the relative similarity structure among phonological features seems to be preserved across positions ($-2, -1, 0, +1, +2$), indicating that each position instantiates a similar phonological subspace.
Together, these findings support our hypothesis that S3Ms encode phonological information in position-dependent, orthogonal subspaces.

Interestingly, positional structure extends beyond immediate neighbors ($\pm1$) to more distant phones ($\pm2$), even though individual frame-level representations exhibit near-zero success rates (\Cref{ss:neighbors}).
This apparent discrepancy suggests that success rate analyses may underestimate the presence of contextual information, as individual frame representations can entangle multiple phonological components, leading to noisy approximations when applying phonological arithmetic at the single frame level.
Consistent with this interpretation, \Cref{fig:layerwise_norm} shows that the norm of phonological vectors decrease with distance from the center phone, following $0 > \pm1 > \pm2$.

\begin{figure}[t!]
    \centering
    \includegraphics[width=0.9\columnwidth]{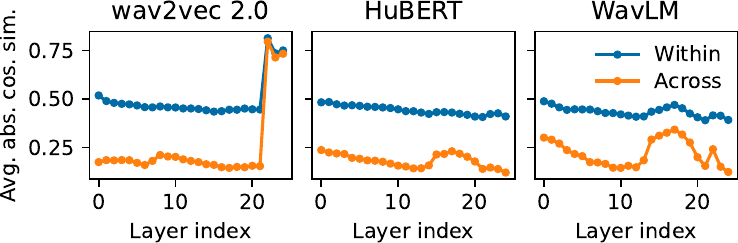}
    \includegraphics[width=0.9\columnwidth]{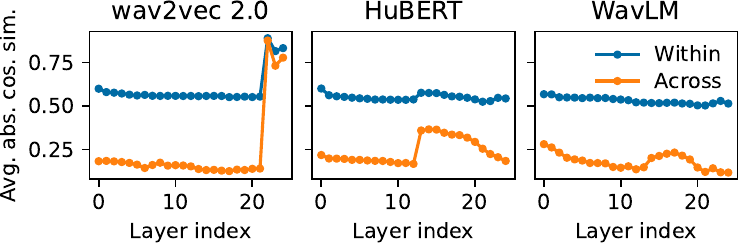}
    \caption{
    Observing the level of orthogonality for each layer. We observe the average of absolute values of cosine similarities within- and across-subspace, on TIMIT (upper) and VoxAngeles (lower).
    Here, we define within- as same natural class (vowel or consonant), and same position ($-2$, $-1$, $0$, $+1$, or $+2$).
    }
    \label{fig:layerwise_orth}
\end{figure}

Finally, we examine whether this positional structure is specific to the final layer of WavLM or persists across models and layers (\Cref{fig:layerwise_orth}).
We compare averaged absolute cosine similarities for \emph{within-position} pairs, constructed by phonological vectors from the same natural class (vowel or consonant) and the same relative position ($-2, -1, 0, +1, +2$), and \emph{across-position} pairs (all remaining combinations).
Across all models and layers, within-position similarities are consistently higher than across-position similarities, implying that positional orthogonality is approximately maintained throughout the layers.

\begin{figure}[!htbp]
    \centering
    \includegraphics[width=0.9\columnwidth]{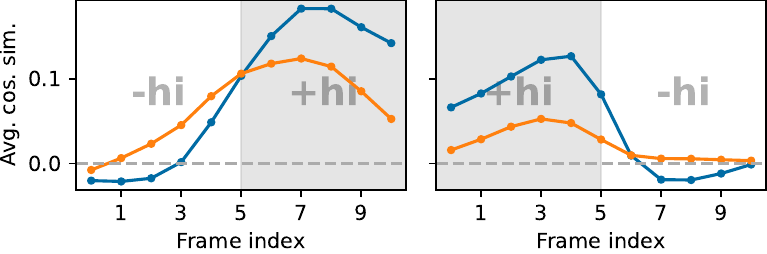}
    \includegraphics[width=0.9\columnwidth]{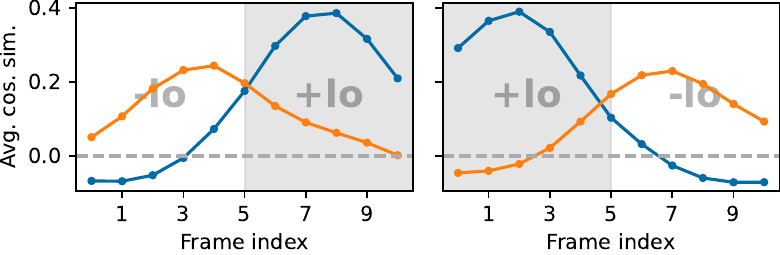}
    \includegraphics[width=0.9\columnwidth]{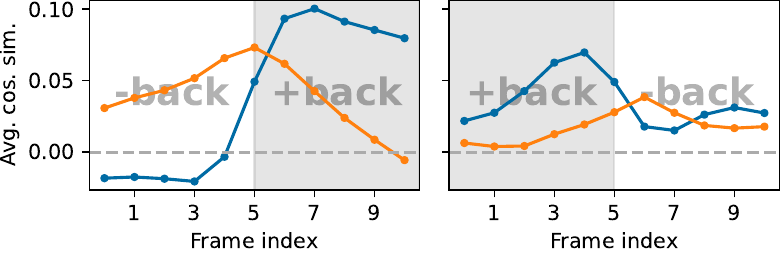}
    \includegraphics[width=0.9\columnwidth]{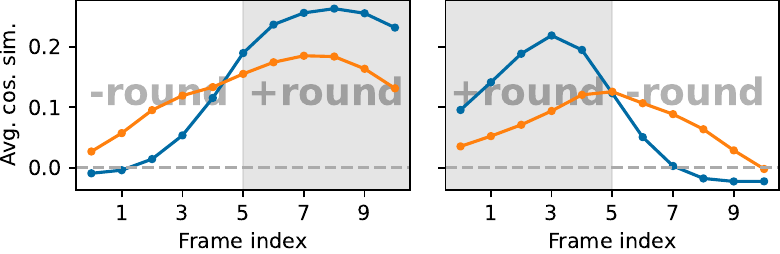}
    \includegraphics[width=0.9\columnwidth]{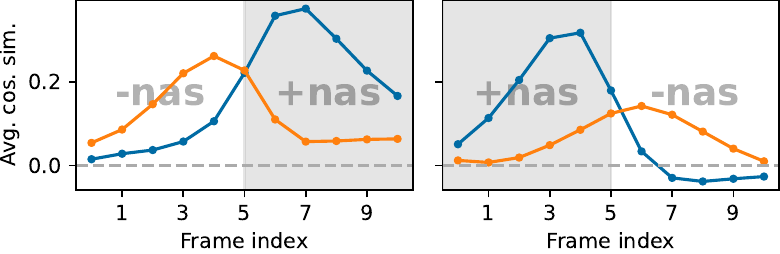}
    \includegraphics[width=0.9\columnwidth]{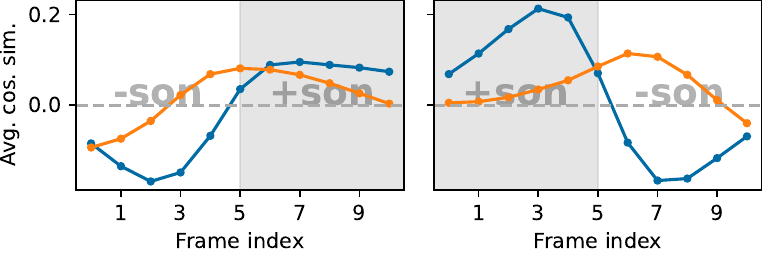}
    \includegraphics[width=0.9\columnwidth]{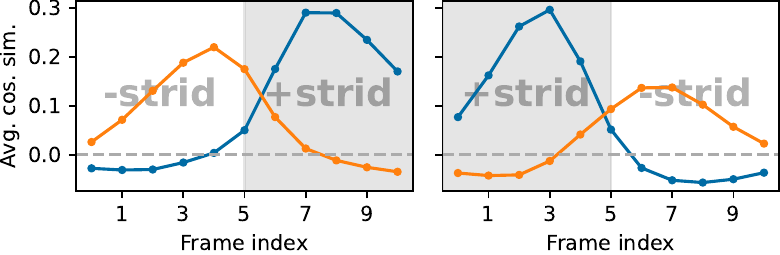}
    \includegraphics[width=0.9\columnwidth]{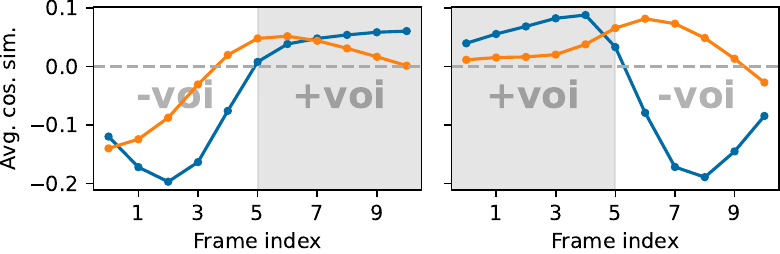}
    \caption{
Phonetic segmentation experiment on TIMIT, where frame index 5 denotes the phonetic boundary of TIMIT.
Cosine similarity between frame-level S3M representations and position-dependent phonological vectors around the phonetic boundaries.
For example, for the boundary between unvoiced \texttt{[-voi]} and voiced \texttt{[+voi]}, left figure (boundary onsets) compares the evidence of current phone being voiced ($\cos(\mathbf{v}^{0}, \mathbf{r}[i])$; blue) and the following phone being voiced ($\cos(\mathbf{v}^{+1}, \mathbf{r}[i])$; orange), whereas the right figure (boundary offsets): compares the evidence of current phone being voiced ($\cos(\mathbf{v}^{0}, \mathbf{r}[i])$; blue) and the preceding phone being voiced ($\cos(\mathbf{v}^{-1}, \mathbf{r}[i])$; orange).
    }
    \label{fig:edges}
\end{figure}

\subsection{Phonetic segmentation}\label{ss:segmentation}

\textbf{Settings.}
The positional orthogonality observed in \Cref{ss:orthogonality} suggests that S3M representations should be sensitive to phonetic boundaries in order to select the appropriate position-dependent phonological subspaces.
If phonological information is encoded relative to previous, current, and next phones, then the model should implicitly track the phonetic boundaries.

To test this hypothesis, we analyze frame-level representations around the phonetic boundaries.
For each boundary, we extract eleven frame-level representations: five frames before the boundary ($\mathbf{r}[0], \ldots, \mathbf{r}[4]$), the boundary frame itself ($\mathbf{r}[5]$), and five frames after the boundary ($\mathbf{r}[6], \ldots, \mathbf{r}[10]$).
For each frame-level representation, we compute cosine similarity with phonological vectors associated with previous ($\mathbf{v}^{-1}$), current ($\mathbf{v}^{0}$), and next phones ($\mathbf{v}^{+1}$).
We consider two types of boundaries: (i) boundary onsets, \textit{i.e.}, boundary between $p^{-1}$ and $p^0$, and (ii) boundary offsets, \textit{i.e.}, boundary between $p^{0}$ and $p^{+1}$.

For example, consider a boundary onset where the previous phone is unvoiced (\texttt{[-voi]}) and the current phone is voiced (\texttt{[+voi]}).
If the model encodes phonological information in a boundary-sensitive manner, the voicing vector associated with the current phone, $\mathbf{v}^{0}_{\text{voi.}}$, should align more with post-boundary frames ($\mathbf{r}[6], \ldots, \mathbf{r}[10]$) than pre-boundary frames ($\mathbf{r}[0], \ldots, \mathbf{r}[4]$).
In other words, we expect $\cos(\mathbf{v}^{0}_{\text{voi.}}, \mathbf{r}[i])$, which shows whether the current phone $p^0$ is voiced, to be higher for post-boundary than pre-boundary frames.

\begin{figure*}[!t]
    \centering
    \includegraphics[width=\textwidth]{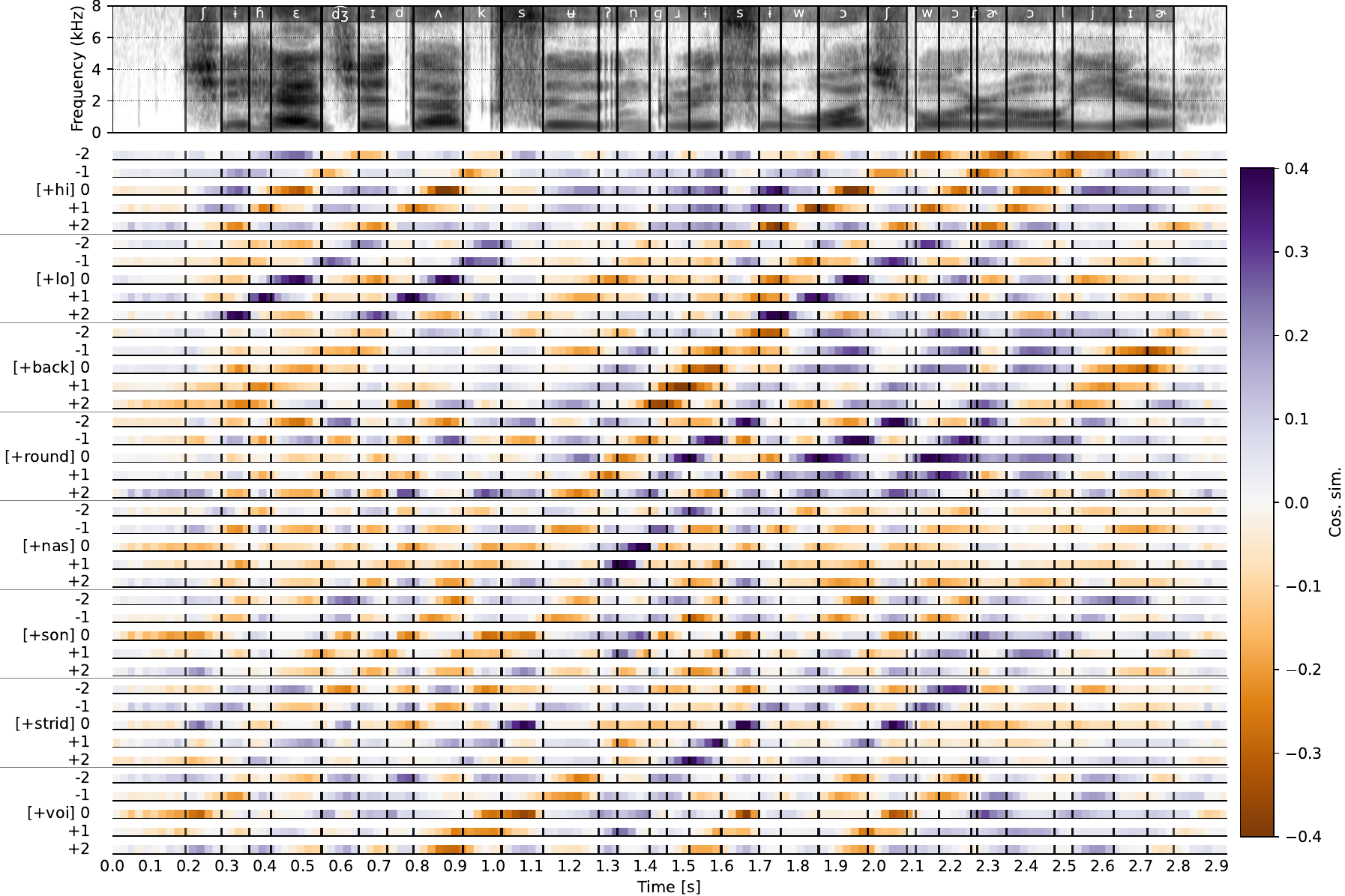}
    \caption{
        Cosine similarity between frame-level S3M representations and phonological vectors for relative phone positions ($-2$ to $+2$) over a full TIMIT utterance.
        Staircase-like transitions often align with phonetic boundaries, illustrating position-specific orthogonal phonological subspaces.
    }
    \label{fig:demo}
\end{figure*}

Under the same example, consider the voicing vector associated with the next phone, $\mathbf{v}^{+1}_{\text{voi.}}$, at the boundary onset.
Because $p^{0}$, \textit{i.e.}, the ``next'' phone of $p^{-1}$, is \texttt{[+voi]}, we expect $\cos(\mathbf{v}^{+1}_{\text{voi.}}, \mathbf{r}[i])$ to be higher for pre-boundary frames (those of $p^{-1}$) than for post-boundary frames (those of $p^0$).
In other words, $\cos(\mathbf{v}^{+1}_{\text{voi.}}, \mathbf{r}[i])$ quantifies the evidence of the ``next'' phone to be voiced.
Similarly, for the voicing vector associated with the previous phone, $\mathbf{v}^{-1}_{\text{voi.}}$, at the boundary offset, $p^{0}$, \textit{i.e.}, the previous phone of $p^{+1}$, is \texttt{[+voi]}.
Accordingly, $\cos(\mathbf{v}^{-1}_{\text{voi.}}, \mathbf{r}[i])$ should be higher for post-boundary frames (those of $p^{+1}$) than for pre-boundary frames (those of $p^0$).

We conduct this analysis on the TIMIT train split using representations from the final layer of WavLM.
Phonetic boundaries are aggregated where the relevant phonological feature changes across the boundary (\textit{e.g.}, \texttt{[-voi]} to \texttt{[+voi]} for boundary onsets, and \texttt{[+voi]} to \texttt{[-voi]} for boundary offsets).
For phonological vectors from vowel features (high, low, back, round), we choose $p^0$ as vowels.
Similarly, for consonantal features (nasal, sonorant, strident, voicing), we choose $p^0$ as consonants.
The reported curves show cosine similarities averaged across all such boundary instances.
Phonological vectors are taken from \Cref{ss:orthogonality}, computed using the TIMIT train split.

\textbf{Results.}
As shown in \Cref{fig:edges}, the frame at which the two cosine similarities intersect coincides with the annotated phonetic boundaries of TIMIT, for both boundary onsets (left) and offsets (right).
These observations suggest that S3Ms segment their frame-level representations approximately according to phonetic boundaries, with position-dependent subspaces constructed based on phonetic segments rather than fixed windows.

\section{Discussions}
\subsection{Qualitative example of phonological vectors}\label{ss:demo}
While \Cref{fig:fig1} illustrated cosine similarities between frame-level S3M representations and phonological vectors over a short segment corresponding to a single word (``catch''), we now extend this visualization to a full TIMIT utterance (``She had your dark suit in greasy wash water all year.'').
Specifically, \Cref{fig:demo} shows cosine similarities with phonological vectors associated with relative phone positions from $-2$ to $+2$ ($p^{-2}$, $p^{-1}$, $p^{0}$, $p^{+1}$, $p^{+2}$).
Although we show eight phonological features used in \Cref{ss:orthogonality,ss:segmentation}, the same visualization can be done to other phonological features provided by \texttt{PanPhon}.

Across the full utterance, we observe a clear staircase-like structure that aligns with phonetic boundaries, rather than seeing a fixed window.
For example, for the vowel \textipa{[E]}, the pattern is particularly evident in the \texttt{[+low]} phonological vector and appears in an inverse form for the \texttt{[+high]} vector, consistent with their phonological opposition.
Although some segments show weaker or less regular transitions, the overall pattern remains clear.

Additionally, staircase-like structure spans relative positions $-2$ through $+2$, consistent with \Cref{ss:orthogonality}.
This visualization further suggests that the success rate-based analysis in \Cref{ss:neighbors} underestimates this effect.

\subsection{Layerwise mask-filling behavior}\label{ss:reconstruction}
\textbf{Settings.}
S3Ms are trained to reconstruct masked portions of the input from surrounding context \cite{schneider2019wav2vec,baevski2020wav2vec,hsu2021hubert,chen2022wavlm}, which encourages representations to encode information about neighboring speech sounds.
Given the coarticulatory nature of speech, this training objective suggests that phonological properties of neighboring phones, \textit{i.e.}, surrounding phonetic environments \cite{choi2025leveraging}, should be leveraged to support accurate mask-filling.

To examine how this context-based prediction behavior varies across layers, we analyze layerwise differences in mask-filling behavior.
Specifically, we compute cosine similarity between representations extracted from the original signal and those extracted from the masked signal, on the masked portion.
Higher similarity indicates more accurate recovery of masked representations from surrounding context.

We use the TIMIT test split and apply each model's default masking configuration (a mask length of 10 frames, approximately 205 ms).
As transformer representations are known to be anisotropic \cite{ethayarajh2019HowCA}, we apply ZCA whitening prior to computing cosine similarity \cite{krizhevsky2009learning}.
Whitening statistics are estimated from 100 randomly sampled utterances from the train split.

\textbf{Results.}
\begin{figure}[t]
    \centering
    \includegraphics[width=1.0\columnwidth]{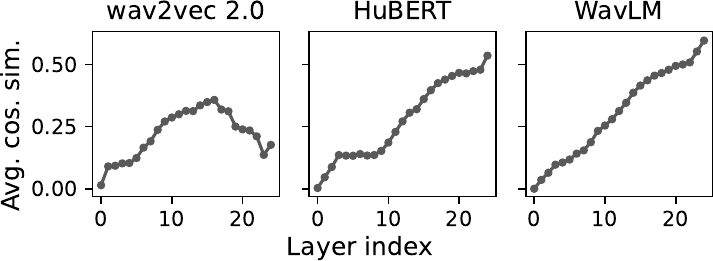}
    \caption{
        Layerwise mask-filling behavior on TIMIT.
        Higher cosine similarity indicates more accurate recovery of masked representations from surrounding context.
    }
    \label{fig:reconstruction}
\end{figure}

As shown in \Cref{fig:reconstruction}, the cosine similarity between representations extracted from the original and masked signals continues to increase across all layers of HuBERT and WavLM.
In contrast, wav2vec 2.0 peaks at an intermediate layer and degrades afterwards.
We suspect that these differences may be related to the layerwise patterns observed in \Cref{ss:pooling} and \Cref{ss:neighbors}, as well as findings from prior work \cite{pasad2021layer,pasad2023comparative}, potentially reflecting differences in training strategies among S3Ms \cite{huo2025iterative}.
Taken together, one possible interpretation of the behavior of HuBERT and WavLM is that all layers are largely driven by mask-filling objectives.
As masking is symmetric, representations at a given position may encode information useful for predicting which phones are likely to occur before or after the current one.
Also, accurate mask-filling might require information about neighboring phones due to coarticulation.
We leave a more concrete investigation to future work.

\subsection{Connection with Observations from Previous Works}
\textbf{Interpretability.} Previous work on S3M interpretability has primarily focused on \emph{what} information is encoded in layerwise representations \cite{pasad2021layer,pasad2023comparative}, such as spectral \cite{choi2022opening,pasad2021layer}, phonetic \cite{choi2024understanding,martin2023probing,choi2024self}, phonemic \cite{martin2023probing}, articulatory \cite{cho2023sslart,cho2024ssluniart}, syllabic \cite{baade2025syllablelm,cho2026sylber,cho2025sylber}, lexical \cite{peng2022word,pasad2024self,choi2024self}, syntactic \cite{shen2023wave}, and semantic information \cite{pasad2024self,choi2024self}.
In contrast, our work focuses on \emph{how} phonetic information is structured and organized within the S3M representations.

\textbf{Phonological vectors.}
\cite{choi2026self} proposed that S3Ms encode phonological information through linear combinations of phonological vectors.
They provide a principled explanation for several empirical observations, including why S3Ms encode phonetic information in terms of relative similarity \cite{choi2022opening,choi2024self}, why discrete units derived from S3Ms often behave similarly to phonetic units \cite{chang2024exploring,baevski2020wav2vec,hsu2021hubert,wells22_interspeech}, and why similarity structures in S3M representations often reflect natural classes \cite{abdullah2023information,sicherman2023analysing}.
We extend the framework by also showing the existence of position-dependent orthogonal phonological subspaces.
This perspective offers a unifying account of prior observations that word-level S3M similarities align closely with phonetic similarity between words \cite{pasad2024self,choi2024self}, and S3Ms encode information about surrounding phonetic environments, \textit{i.e.}, allophones \cite{choi2025leveraging}.

\textbf{Context-dependent triphone HMMs.}
Before deep neural networks, automatic speech recognition systems often modeled coarticulation using context-dependent models \cite{schwartz1985context,young1994tree}.
Our results suggest that S3Ms may rediscover similar contextual structure through position-dependent phonological vectors.

\textbf{Effective context window.}
The results in \Cref{ss:neighbors} connect to prior analyses of the effective temporal context size of S3Ms \cite{meng25b_interspeech,choi2025device}.
Whereas earlier work often characterized contextualization in terms of fixed temporal windows, our results suggest that S3M representations are organized with respect to phonetic boundaries, while position-dependent subspaces are weighted per their proximity to the center phone (\Cref{fig:layerwise_norm}).

\textbf{Unsupervised segmentation.}
Phonetic segmentation analysis in \Cref{ss:segmentation} sheds new light on how word- and syllable-like boundaries can emerge without supervision \cite{pasad2024self,baade2025syllablelm,cho2025sylber,cho2026sylber,visser2026zerosyl,li2023dissecting}.
By demonstrating that S3M representations implicitly encode position-dependent structure aligned with phonetic boundaries, we identify a concrete signal that can be exploited for unsupervised phone- and syllable-level segmentation.

\subsection{Implications for Future Work}
\textbf{Interpretability.}
Our findings closely relate to \cite{liu2023self,feng2022silence,kamper2025linearvc}, which suggests that speaker information in S3Ms is encoded in orthogonal subspaces and can be characterized by linear vectors.
Together with our results, this points to a broader representational principle in S3Ms: different sources of information (\textit{e.g.}, phonological, speaker, or acoustic) might be organized into structured, disentangled subspaces.
An important direction for future work is to characterize how these are organized and interact within the representation space.

\textbf{Discrete speech units.}
Our results point to possible differences between discrete units extracted from S3Ms \cite{chang2024exploring} and acoustic tokens learned by neural audio codecs \cite{mousavi2025discrete}.
It is known that discrete units from S3Ms naturally exhibit greater structural coherence \cite{borsos2023audiolm} and better ASR performance \cite{chang2024exploring} than acoustic tokens.
We hypothesize that this difference arises from encoding phonetic context.
One promising direction can be on novel architectures and training objectives aimed at disentangling context-independent and context-dependent representations \cite{defossez2024moshi,zhang2024speechtokenizer,huang2026kanade}.
Further, \Cref{ss:orthogonality} implies that phonological structure is similar for multiple relative positions, raising the possibility of designing more structured neural architectures that explicitly model this property.

\textbf{Interpretable speech representations.}
The observed phonological structure in \Cref{ss:demo} connects naturally to phonetic and phonological posteriograms \cite{hazen2009query,morrison2024fine,cernak2015phonological} and articulatory inversion \cite{cho2024coding}.
These connections suggest that representation similar to \Cref{fig:demo} can serve as interpretable and linguistically grounded intermediate representations for applications such as vocoders and text-to-speech models.

\section{Conclusion}
In this work, we investigate how S3Ms contextualize at the level of individual frame representations.
Through a series of analyses, we showed that S3Ms encode phonological information of phone sequences in a compositional manner.
Individual frames reflect properties of the current and neighboring phones via phonological vectors.
Further, we showed that such vectors lies within relative position-dependent orthogonal subspaces.
Finally, we demonstrated that this structure leads to emergent phonetic boundaries.
Together, these findings provide a unified understanding of how transformer-based context networks of S3Ms encode phonetic context.

\ifcameraready
\section{Acknowledgments}
The authors would like to thank Stephen McIntosh and Ian Shih for their valuable feedback.
\fi

\section{Generative AI Use Disclosure}
Generative AI tools were used during the preparation of this manuscript, primarily for code auto-completion and minor editing related to wording and grammar.
All scientific content and code implementations were independently developed, verified, and finalized by the authors.

\bibliographystyle{IEEEtran}
\bibliography{mybib}

\end{document}